# *Saiyan*: Design and Implementation of a Low-power Demodulator for LoRa Backscatter Systems

Xiuzhen Guo
*Tsinghua University*

Longfei Shangguan
*University of Pittsburgh & Microsoft*

Yuan He
*Tsinghua University*

Nan Jing
*Yanshan University*

Jiacheng Zhang
*Tsinghua University*

Haotian Jiang
*Tsinghua University*

Yunhao Liu
*Tsinghua University*

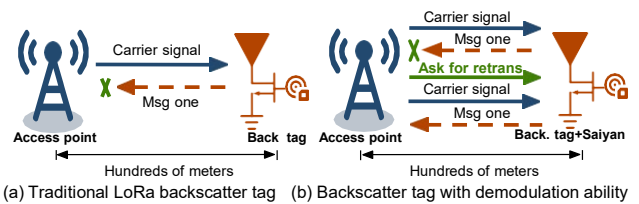

Figure 1: Saiyan empowers the LoRa backscatter tag to demodulate feedback signals from a remote access point.

## Abstract

The radio range of backscatter systems continues growing as new wireless communication primitives are continuously invented. Nevertheless, both the bit error rate and the packet loss rate of backscatter signals increase rapidly with the radio range, thereby necessitating the cooperation between the access point and the backscatter tags through a feedback loop. Unfortunately, the low-power nature of backscatter tags limits their ability to demodulate feedback signals from a remote access point and scales down to such circumstances.

This paper presents Saiyan, an ultra-low-power demodulator for long-range LoRa backscatter systems. With Saiyan, a backscatter tag can demodulate feedback signals from a remote access point with moderate power consumption and then perform an immediate packet re-transmission in the presence of packet loss. Moreover, Saiyan enables rate adaption and channel hopping – two PHY-layer operations that are important to channel efficiency yet unavailable on long-range backscatter systems. We prototype Saiyan on a two-layer PCB board and evaluate its performance in different environments. Results show that Saiyan achieves 3.5–5× gain on the demodulation range, compared with state-of-the-art systems. Our ASIC simulation shows that the power consumption of Saiyan is around 93.2 $\mu W$ . Code and hardware schematics can be found at: https://github.com/ZangJac/Saiyan.

## 1 Introduction

Backscatter radios have emerged as an ultra-low-power and economical alternative to active radios. The ability to communicate over long distances is critical to the practical deployment of backscatter systems, particularly in outdoor scenarios (*e.g.*, smart farm) where backscatter tags need to deliver their data to a remote access point regularly. Conventional RFID technology [12] functions within only a few meters and is not well suited for outdoor scenarios. To this end, the research community has proposed long-range backscatter approaches [23, 40, 47] that leverage Chirp Spreading Spectrum (CSS) modulation on LoRa [5] to improve the signal resilience to noise, thereby extending the communication range. For instance, LoRa backscatter [47] allows tags to communicate with a source and a receiver separated by 475 m.

However, existing long-range LoRa backscatter systems present a new challenge on packet delivery. The backscatter signals travel twice the link distance and suffer drastic attenuation as the link distance scales. They become very weak after traveling long distances, thereby causing severe bit errors and packet losses. Figure 2 shows the Bit Error Rate (BER) of PLoRa [40] and Aloba [23], two representative long-range LoRa backscatter systems. Evidently, the BER of both systems rises rapidly from less than 1% to over 50% as the tag is moved away from the transmitter (Tx). The receiver is almost unable to demodulate any backscatter signal once the tag is placed 20 m away from the transmitter. Considering that the backscatter tags are unaware of packet loss, each packet must be transmitted blindly for multiple times to lift the packet delivery ratio, which inevitably wastes precious energy and wireless spectrum and cause interference to other radios that work on the same frequency band.

To address these issues, we expect a *downlink* from the access point to the backscatter tag, through which the feedback signals (*e.g.*, asking for a packet re-transmission) can be delivered, thereby forming a *feedback loop*. We envision that such a feedback loop will bring opportunities to bridge the gap between long-range backscatter communication and the growing packet loss rate, as reflected on the following aspects:

. *Making on-demand re-transmissions in the presence of packet loss*. The backscatter tag demodulates feedback signals from an access point and makes a re-transmission only if it is asked to do so. This reactive packet re-transmission can mitigate packet loss while improving power and channel efficiency.

. *Scheduling channel hopping to minimize interference*. The unlicensed band where the LoRa resides in is already over crowded. The access point monitors the wireless spectrum and notifies the backscatter tag to switch channels in the presence of in-band interference. As such, the channel utilization and packet delivery ratio can be improved effectively.

. *Adapting data rate to link condition*. The condition of backscatter links varies over time. The access point assesses each backscatter link and keeps the backscatter tag updated through the feedback loop. Each tag then adapts its data rate proactively to utilize the wireless link better.

In addition, such a feedback loop empowers the network administrator to turn on/off sensors on backscatter tags remotely, thereby avoiding labor-intensive and time-consuming physical access to the devices.

To enjoy these benefits, the primary hurdle to overcome is the packet demodulation on LoRa backscatter tags. LoRa is based on frequency modulation. To demodulate a LoRa symbol, the commercial LoRa receiver operates by down-converting the incident LoRa chirp to the baseband, sampling it at twice the chirp bandwidth (BW), and then converting the signal samples from the time domain to the frequency domain using Fast Fourier Transformation (FFT). These operations consume over 40 $mW$ power altogether [6]. Considering a miniaturized energy harvester equipped with a palm-sized solar panel, this harvester merely generates 1 $mW$ power every 25.4 seconds in a bright day [3]. In other words, to support the standard LoRa demodulation, a backscatter tag needs to wait for 17 minutes until it accumulates enough power. Although the envelope detector has been used for packet demodulation on many backscatter systems [38, 46, 52], it is ill-suited for LoRa demodulation because the envelope of a LoRa signal is a constant.

In this paper, we propose Saiyan, a low-power demodulator for long-range LoRa backscatter systems. Saiyan is based on an observation that *a frequency-modulated chirp signal can be transformed into an amplitude-modulated signal using a differential circuit*. The amplitude of this transformed signal scales with the frequency of the incident chirp signal, thereby allowing us to demodulate a LoRa chirp by tracking the peak amplitude on its transformed counterpart without using power-intensive hardware, such as a down-converter and an ADC. To put this high-level idea into practice, the challenges in design

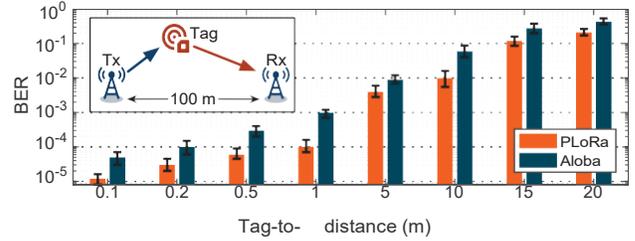

Figure 2: BER of PLoRa [40] and Aloba [23] in different tag-to-transmitter settings. The BER of both systems rises dramatically with the increasing distance between the transmitter and the tag. We re-implement both systems on PCB.

and implementation must be addressed, as summarized below.

**Frequency-amplitude transformation**. The low-power nature of backscatter tags requires the differential circuit to be extremely low-power. Moreover, to support higher data rate, such a differential circuit should also be hyper-sensitive to the frequency variation of LoRa signals. However, the narrow bandwidth of LoRa signals (*e.g.*, 125/250/500 KHz) renders the conventional detuning circuits, such as RLC resonant circuit, inapplicable. In Saiyan, we instead repurpose the Surface Acoustic Wave (SAW) filter as a signal converter by leveraging its sharp frequency response (§2.1). To minimize the power consumption on demodulation, we further replace the power-consuming ADC with a well-designed double-threshold based comparator (§2.2) coupled by a proactive voltage sampler (§2.3).

**Improving the demodulation sensitivity**. Although the aforementioned vanilla Saiyan can demodulate LoRa signals with the minimum power consumption, its communication range is limited to 55 m because of the Signal-to-Noise Ratio (SNR) losses in both SAW filter and envelope detector. To extend the communication range, we introduce a low-power cyclic-frequency shifting circuit coupled with an Intermediate Frequency (IF) amplifier to simultaneously remove the noise while magnifying the signal power. This low-power circuit brings 11 dB SNR gain and doubles the demodulation range (§3.1). Furthermore, a low-power correlator is leveraged to extend the demodulation range further to 148 m (§3.2).

**Implementation**. We implement Saiyan on a 25 $mm \times$ 20 $mm$ two-layer Printed Circuit Board (PCB) using analog circuit components and an ultra-low power Apollo2 MCU [13]. The Application Specific Integrated Circuit (ASIC) simulation shows that the power consumption can be reduced to 93.2 $\mu$W, which is affordable on an energy harvesting tag. The main contributions of this paper are summarized as follows:
. We simplify the standard LoRa demodulation from energy perspective and design the first-of-its-kind low-power LoRa demodulator that can run on an energy harvesting tag.
. We design a set of simple but effective circuits and algorithms, prototyping them on PCB board for system evaluation.

. We demonstrate that Saiyan outperforms the state-of-the-arts on power consumption, communication range, and throughput.

The remainder of this paper is structured as follows. We present the design of vanilla Saiyan in Section 2, followed by super Saiyan in Section 3. Section 4 describes the implementation details. The experiment (§5) follows. We review related works in Section 6 and conclude in Section 7.

## 2 Vanilla Saiyan

A LoRa symbol is represented by a chirp whose frequency grows linearly over time, as formulated below.

$$s(t) = A sin(2\pi f(t)t) \quad (1)$$

where $A$ is the signal amplitude; $f(t) = F_0 + kt$ describes how the frequency of this chirp signal varies over time; $F_0$ is the initial frequency offset; $k$ is the frequency changing rate. The frequency of a LoRa chirp wraps to 0 right after peaking $BW$ —the bandwidth of LoRa. Different LoRa chirps peak the frequency $BW$ at different time due to the difference in their initial frequency offset, as shown in Figure 3(a). Applying a differential to a LoRa chirp, we have:

$$s'(t) = \frac{ds(t)}{dt} = Acos(2\pi f(t)t)[2\pi \frac{df(t)}{dt}t + 2\pi f(t)] \quad (2)$$
$$= 2\pi \underbrace{A(F_0 + 2kt)}_{Amplitude} cos(2\pi f(t)t)$$

The above equation indicates that the amplitude of the transformed signal $s'(t)$ is proportional to the frequency of the input LoRa chirp $s(t)$, as shown in Figure 3(b). This frequency-amplitude correlation allows us to demodulate the frequency-modulated (FM) chirp signal by tracking the peak amplitude of its transformed amplitude-modulated (AM) counterpart.

### 2.1 Frequency-amplitude Transformation

To realize the differential operation [10], an intuitive solution is using RLC resonant circuit. However, the narrow bandwidth of LoRa (*e.g.*, 125/250/500 KHz) renders this idea infeasible (see Appendix A.1 for details). In Saiyan we instead exploit the sharp frequency response of the Surface Acoustic Wave (SAW) filter to transform LoRa chirps into amplitude-modulated signals.

**SAW filter primer**. A SAW filter consists of two interdigital transducers (shown in Figure 4). The input interdigital transducer transforms electrical signals into acoustic waves; the output interdigital transducer then transforms acoustic waves back into electrical signals. This two-stage signal transformation introduces 6 dB insertion loss to the incident signal [4].

**Re-purposing SAW filter as a signal converter**. Our design is based on the observation that the frequency response of a SAW filter grows monotonically within a certain frequency band (termed as *critical band*). After passing through the critical band, the chirp signal will be transformed into an AM signal whose amplitude scales with the frequency of this input FM chirp. This allows us to demodulate LoRa chirp by simply tracking the peak amplitude of the AM signal. On the other hand, since SAW filter by its own design is battery-free, such frequency-amplitude transformation doesn't incur extra power consumption to backscatter tags.

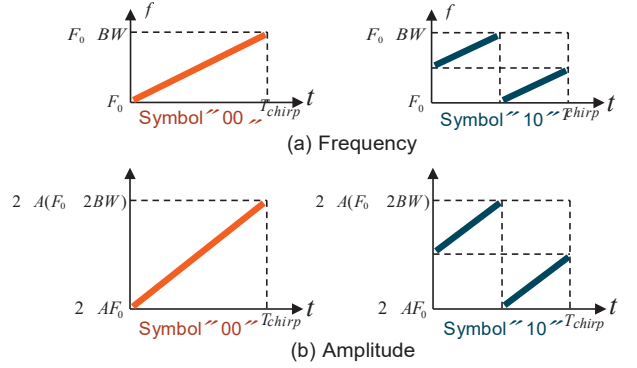

Figure 3: LoRa symbols before and after frequency-amplitude transformation. (a) Different LoRa symbols in the frequency domain. (b) The amplitude of LoRa symbols after frequency-amplitude transformation.

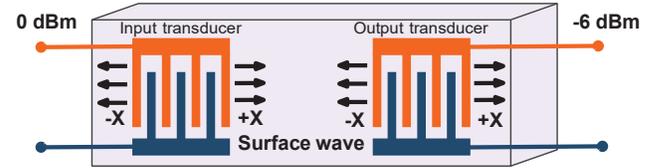

Figure 4: Diagram of the SAW filter. The SAW filter converts electrical signal into acoustic signal and then back through two inter-digital transducers.

In Saiyan, we take into account the working frequency and bandwidth of LoRa signals and select a general-purpose Qualcomm B3790 [1] SAW filter as the signal converter. Figure 5 shows its frequency response. The signal amplitude grows by 25 dB as the frequency of the incident signal scales from 433.5 MHz to 434 MHz (500 KHz bandwidth). To validate this 25 dB amplitude gap is strong enough for differentiating LoRa chirps, we feed four different chirp symbols into this SAW filter and plot the output in Figure 6. Evidently, these symbols peak their amplitude at clearly different time points, confirming the effectiveness of the SAW filter.

### 2.2 Demodulation

The transformed symbols are down-converted to the baseband through an envelope detector. Before demodulation, the standard LoRa receiver first digitizes these baseband signals using an ADC, which is power intensive. To save power, an intuitive solution is to replace the ADC with a low-power voltage comparator [37, 46]. The threshold of this comparator is set to a value slightly lower than the peak amplitude of the basband signal. This allows us to locate the peak amplitude by checking the comparator's output. Unfortunately, due to the in-band interference and hardware noise, the transformed AM signal may experience multiple amplitude peaks and valleys that may confuse the comparator.

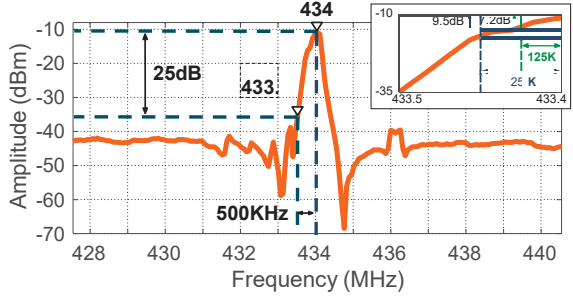

Figure 5: The amplitude-frequency response of the SAW filter adopted by Saiyan. The central frequency is 434 MHz. The measured insertion loss of this SAW filter is 10 dB. We observe 25 dB, 9.5 dB, and 7.2 dB amplitude variation as the frequency of an incident signal grows from 433.5 MHz, 433.75 MHz, and 433.875 MHz to 434 MHz, respectively.

Table 1: The required sampling rate (KHz) in theory/practice to achieve 99.9% decoding accuracy.

|   | SF=7 | SF=8 | SF=9 | SF=10 | SF=11 | SF=12 |
|---|------|------|------|-------|-------|-------|
| **K=1** | 15.6/20 | 7.8/12 | 3.9/5.5 | 1.95/2.6 | 0.98/1.2 | 0.49/0.6 |
| **K=2** | 31.2/40 | 15.6/20 | 7.8/12 | 3.9/5.5 | 1.95/2.6 | 0.98/1.2 |
| **K=3** | 62.5/85 | 31.2/40 | 15.6/20 | 7.8/12 | 3.9/5.5 | 1.95/2.6 |
| **K=4** | 125/180 | 62.5/85 | 31.2/40 | 15.6/20 | 7.8/12 | 3.9/5.5 |
| **K=5** | 250/400 | 125/180 | 62.5/85 | 31.2/40 | 15.6/20 | 7.8/12 |

**Double-threshold based comparator**. In Saiyan we instead adopt a double-threshold based comparator to stabilize the output binary sequence. Let $U_H$ and $U_L$ denote the high-voltage and low-voltage threshold defined in this comparator. When the amplitude of an input signal is sufficiently higher than $U_H$, the comparator outputs a high voltage. When the amplitude of this signal is equivalent to $U_H$ or above, no chattering occurs since the output will not respond unless the input falls below $U_L$. Following this idea, the output voltage $B_i$ can be formulated as follows:

$$B_i = \begin{cases} low, & if\ A_i < U_H\ \&\ B_{i-1} = low \\ high, & if\ A_i \geq U_H\ \&\ B_{i-1} = low \\ low, & if\ A_i < U_L\ \&\ B_{i-1} = high \\ high, & if\ A_i \geq U_L\ \&\ B_{i-1} = high \end{cases} \quad (3)$$

where $A_i$ represents the amplitude of the $i^{th}$ signal sample. This double-threshold comparator takes into account the amplitude samples both in the past and present. It nulls out the chattering caused by the amplitude oscillation. The threshold setup is detailed in system implementation (§4).

To show the effectiveness of this double-threshold based comparator, we apply it to a LoRa chirp and plot the output in Figure 7. For comparison, we also plot the output of two single-threshold based comparators ($U_H$ alone and $U_L$ alone, respectively). We can see that using a high threshold $U_H$ alone fails to detect the amplitude peak due to the valleys emerging on signal amplitude (i.e., $t \in [t_E, t_F]$ in Figure 7(b)). Using a low threshold $U_L$ alone causes false positives due to the misleading peak emerging on the signal amplitude ($t \in [t_A, t_B]$ in Figure 7(d)). In contrast, the double-threshold based comparator produces a series of stable binary voltages that can guide us to locate the peak amplitude at the correct position (i.e., at the tail of the high voltage samples $t_F$ shown in Figure 7(e)).

### 2.3 Low-power Voltage Sampler

The comparator quantizes chirp samples into binary voltages which are stored in a counter of micro-controller (MCU). The sampling rate tradeoffs the power consumption and the demodulation performance and thus cannot be set arbitrarily. A higher sampling rate supports a higher link throughput. It however consumes more power. Suppose a LoRa chirp encodes $K$ bits data. The data rate equals $K \cdot BW/2^{SF}$, where $BW$ and $SF$ respectively represent bandwidth and spreading factor. According to the Nyquist sampling theorem, the sampling rate should be not lower than $2 \cdot BW/2^{SF-K}$.

However, in reality, using the theoretical minimum sampling rate exacerbates bit errors. We conduct a benchmark experiment to measure the practical sampling rate required to achieve 99.9% decoding accuracy. Table 1 lists the results with different settings of spreading factor and coding rate. We find that the required sampling rate in practice is slightly higher than the theoretical minimum sampling rate. Suggested by this result, we conservatively set the sampling rate to $3.2 \cdot BW/2^{SF-K}$, which guarantees the demodulation performance.

**Decoding**. After quantization, the low-power MCU decodes each LoRa chirp by localizing the bit '1' within each LoRa symbol, as shown in Figure 8. The LoRa preamble contains ten identical up-chirps. Upon detecting the LoRa preamble, Saiyan waits for 2.25 symbol times (sync. symbols) and operates demodulation on the payload hereafter.

**Remarks**. The vanilla Saiyan demodulates LoRa signals with the minimum power consumption. However, its demodulation sensitivity is limited due to the signal attenuation in the SAW filter and the noise added by the envelope detector. Next, we introduce super Saiyan to improve the sensitivity.

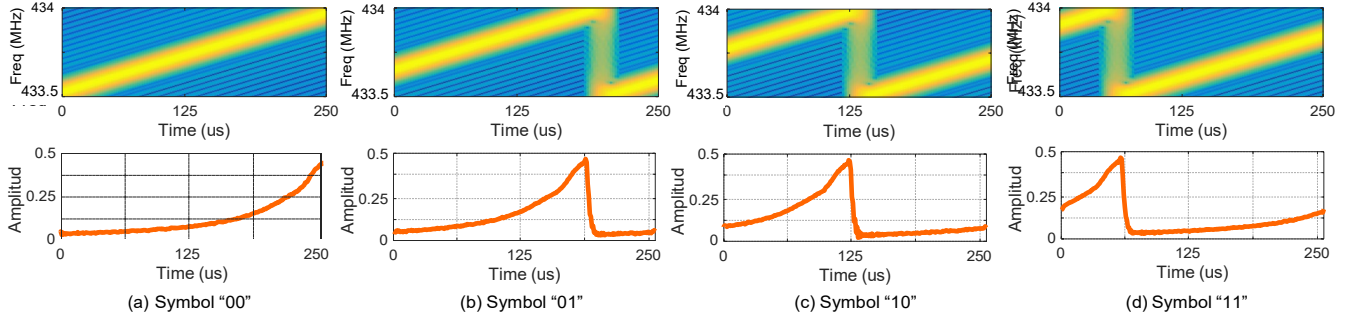

Figure 6: The input (top) and output (bottom) signals of the SAW filter. The amplitude of the output signal scales with the frequency of the input signal. They reach the maximal value simultaneously.

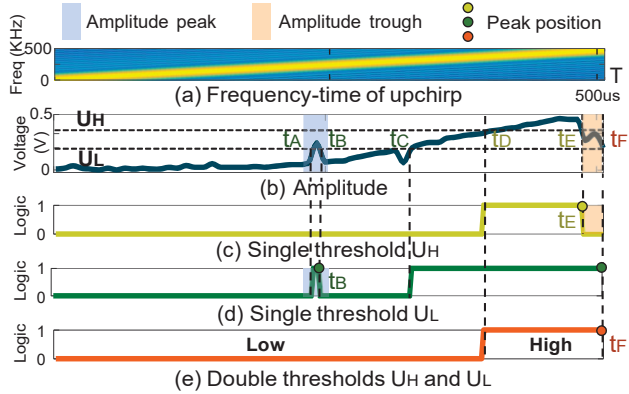

Figure 7: Comparing the output of different voltage comparators. (a): the incident LoRa chirp. (b): the output of an envelope detector. (c)-(d): the output of the single-threshold based comparator that uses $U_H$ or $U_L$ as the cut-off amplitude. (e) the output of the double-threshold based comparator that uses $U_H$ and $U_L$ simultaneously as the cut-off amplitudes.

## 3 Super Saiyan

Super Saiyan takes the following actions to consistently improve the demodulation sensitivity: *i*) improving the SNR of baseband chirp signals with a cyclic-frequency shifting circuit, and *ii*) improving the sensitivity of demodulator with correlation.

### 3.1 Cyclic-frequency Shifting

**Understanding the principle of envelope detector**. The envelope detector has been widely adopted by low-power RF devices to down-convert the incident signal. However, due to the inherent non-linearity caused by the squaring operation of CMOS devices [27], both the targeted signal (*i.e.*, feedback signals from the LoRa access point) and the RF noises will be down-converted to the baseband. Consequently, the targeted signal becomes even weaker after down-conversion. We explicate this phenomenon using the following example. Let $S_{in}$ be the incident signal: $S_{in} = S_t + S_n$, where $S_t$ and $S_n$ denote the targeted signal and RF noises, respectively. The output signal $S_{out}$ of this envelope detector can be represented by:

$$S_{out} = kS_i^2 = k(S_t + S_n)^2 \\ = kS_t^2 + 2kS_t \cdot S_n + kS_n^2 \quad (4)$$

where $k$ represents the attenuation factor. The first term $S_t^2$ on the right side of this equation manifests that the targeted signal $S_t$ is shifted to the baseband through self-mixing. The second and the third terms both indicate the RF noises are shifted to the baseband after mixed with the targeted signal and the noises themselves, respectively, causing strong interference on the baseband.

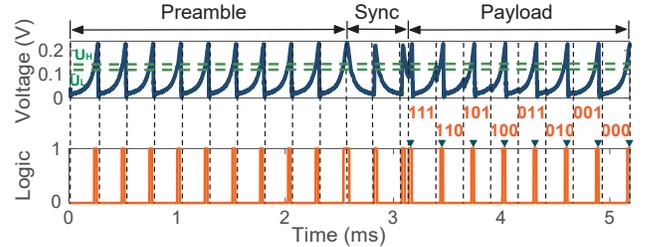

Figure 8: The decoding process of a LoRa packet

**Cyclic-frequency shifting**. In Saiyan we design a low-power circuit to mitigate the SNR loss brought by the envelope detector. The circuit is realized by two RF mixers and two clock signals. Its operation is detailed as follows.

. **Step 1**. The micro-controller first generates a clock signal $CLK_{in}(\Delta f)$ and mixes it with the incident signal $S(F)$, resulting in two sideband signals $S(F - \Delta f)$ and $S(F + \Delta f)$, as shown in Figure 9(a)-(b). The sideband signals and the incident signal are then down-converted to the intermediate frequency (IF) band (denoted by $S(-\Delta f)$ and $S(\Delta f)$) and the baseband (denoted by $S(0)$) respectively with an envelope detector (Figure 9(c)).

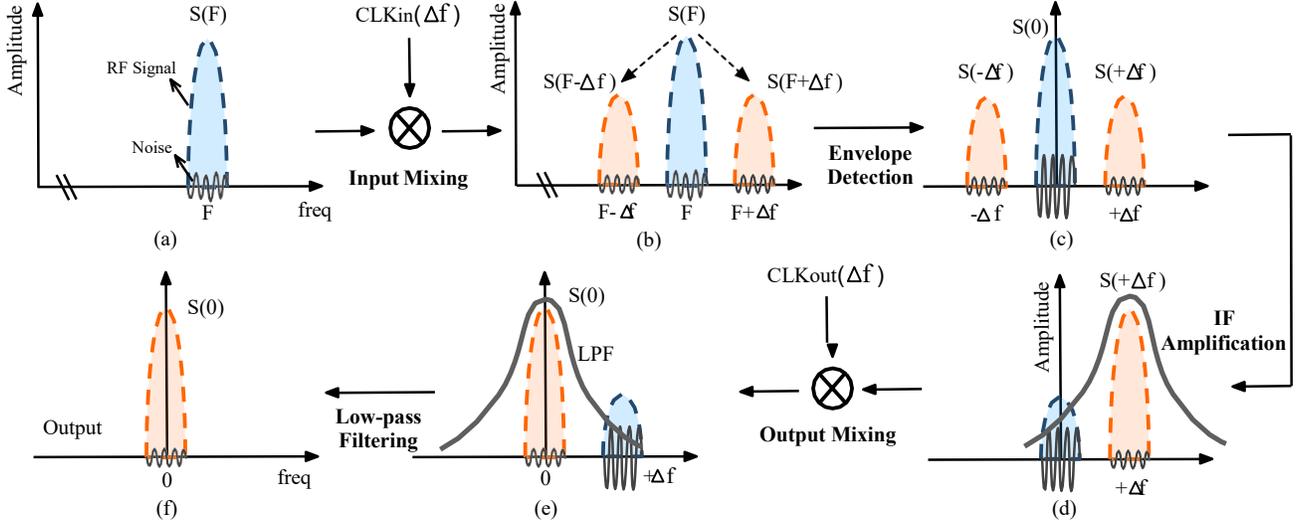

Figure 9: The illustration of the cyclic-frequency shifting. (a) The input signal $S(F)$. (b) $S(F)$ is first mixed with the clock signal, resulting in two sideband signals $S(F-\Delta f)$ and $S(F+\Delta f)$. (c) The envelope detector extracts the envelope of those three signals and down-converts them to the the baseband. (d) The IF amplifier boosts the power of $S(\Delta f)$ and attenuates the power at other frequency bands. (e) The desired signal $S(\Delta f)$ with significantly lower noises is shifted back to the baseband. (f) The output signal $S(0)$.

**. Step 2**. Since RF noises are not down-converted to the IF band by the envelope detector, we amplify the unpolluted IF signal $S(\Delta f)$ using a low-power IF amplifier. The frequency selectivity of this IF amplifier filters out signals at other frequencies (*e.g.*, $S(0)$), as shown in Figure 9(d).

**. Step 3**. The power-amplified IF signal $S(\Delta f)$, mixed with another clock signal $CLK_{out}(\Delta f)$, is shifted back to the baseband, as shown in Figure 9(e). At the same time, the noisy baseband signal $S(0)$ will be shifted to the IF band and then filtered by a low-pass filter (Figure 9(f)).

In a nutshell, this circuit first moves the targeted signal to an intermittent frequency band (step 1) to avoid the RF noise contamination introduced by the envelope detector. This also leaves us an opportunity to remedy the SNR loss in down-conversion (step 2). Finally, the targeted signal is moved back to the baseband for demodulation. At the same time the DC offset, flicker and other noises are moved to the IF band and removed by a low-pass filter (step 3).

Figure 10 shows the spectrums before and after feeding the chirp signal into the cyclic frequency shifting circuit. Evidently, both the inband and out-of-band RF noises have been cleaned by the circuit, ensuring the decodability of chirp signals. Our quantitative measurement shows that the cyclic-frequency shifting circuit brings in 11 dB SNR gain.

**Clock signal generation**. The above circuit design relies on two clock signals $CLK_{in}(\Delta f)$ and $CLK_{out}(\Delta f)$. To save power, we program the MCU to generate $CLK_{in}(\Delta f)$ signal and then leverage a delay line to copy $CLK_{in}(\Delta f)$ as $CLK_{out}(\Delta f)$:

$$CLK_{out}(\Delta f) = CLK_{in}(\Delta f + \Delta \phi) \quad (5)$$

where $\Delta \phi$ is the phase shift caused by the delay line. We tune the length of this delay line to ensure $cos(\Delta \phi) \gtrsim 1$ so that $CLK_{out}(\Delta f)$ equals $CLK_{in}(\Delta f)$.

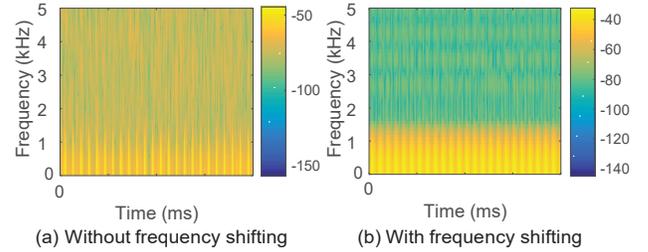

Figure 10: The spectrum of an incident LoRa signal when being down-converted into the baseband with an envelope detector. (a) Without cyclic-frequency shifting. (b) With cyclic-frequency shifting. The LoRa signal contains 24 LoRa chirps ($BW$=500KHz, $SF$=8).

**Circuit integration**. We integrate this cyclic-frequency shifting circuit into the envelope detector. Figure 11 shows the schematic of this design. It consists of an input mixer, an output mixer, an envelope detector, an IF amplifier, a low-pass filter (LPF), an oscillator, and a transmission line. Specifically, The base clock signal is provided by a micro-power precision oscillator LTC6907 [11]. A low-power transistor 2N222 [8] is adopted as the IF amplifier.

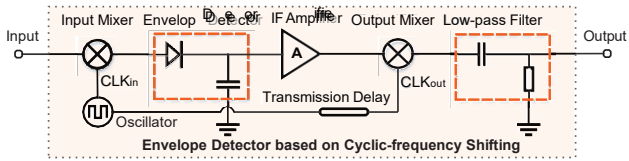

Figure 11: The schematic of cyclic-frequency shifting.

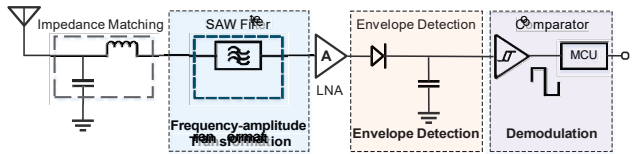

Figure 12: The high-level circuit schematic of Saiyan.

## 3.2 Correlation

While the above cyclic-frequency shifting circuit successfully improves the SNR of the incident signal, the demodulation accuracy still suffers degradation when the incident signal is too weak, *e.g.*, close to the noise floor. We thus employ correlation — a mainstream approach that has been largely adopted for packet detection to further improve the demodulation sensitivity. It operates by correlating signals samples with a local chirp template. An energy peak shows up as long as the incident signal matches the template. The receiver then tracks the energy peak and demodulates the incident signal.

## 4 Implementation

We describe the system implementation in this section.

### 4.1 Backscatter Tag

We implement Saiyan on a 25 $mm \times$ 20 $mm$ two-layer PCB using commercial off-the-shelf analog components and an ultra-low power Apollo2 (10 $\mu$A/MHz) [13] MCU. We determine its size through a mixed analytical and experimental approach, striking a balance between the form factor and circuit interference. Figure 13 shows the hardware prototype. Saiyan functions with an omni-directional antenna [2] with 3 dBi gain.

**Architecture and workflow**. Figure 12 shows the architecture of Saiyan. The incident signal passes through a passive SAW chip B39431B3790Z810 [1] and is transformed into an amplitude-modulated signal. We place a common-gate low-noise amplifier (CGLNA) [17] between the SAW filter and the customized envelope detector to amplify the transformed signal. The amplified signal is then down-converted to the baseband through the envelope detector. Finally, a low-power voltage comparator NCS2202 [9] is leveraged to quantize the output signal from the envelope detector.

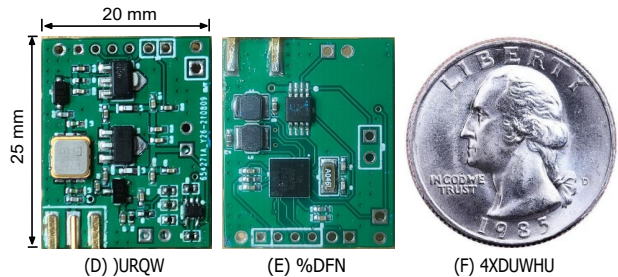

Figure 13: The hardware prototype of Saiyan. The quarter next to Saiyan demonstrates the form factor.

**Plug-and-play**. As an ultra-low-power peripheral, Saiyan can be integrated into the existing long-range LoRa backscatter systems [23, 40] with ignorable engineering efforts. Taking PLoRa [40] as an example, we replace its packet detection module with Saiyan and retained all the remaining functional units the same. This simple replacement allows PLoRa tag to demodulate the feedback signals while retaining the modulation capability at the same time. On the software side, we replicate the sampling rate control logic to facilitate the demodulation.

**Power management**. The energy harvester on Saiyan comprises of a palm-sized photovoltaic panel and a high-efficiency step-up DC/DC converter LTC3105 [3]. It generates 1 $mW$ power every 25.4 seconds in a bright day. The power management module provides a constant 3.3V output voltage to the MCU. The power consumption of this power management module in working mode is approximately 24 $\mu W$

**Determining the voltage thresholds** $U_H$ **and** $U_L$. Ideally, $U_H$ should be slightly lower than the peak amplitude of the input signal $A_{max}$. Let $G$ be the gap between $A_{max}$ and the voltage threshold $U_H$. We have: $G = 20lg(A_{max}/U_H)$. Thus, $U_H$ can be estimated on the basis of the following equation: $U_H = A_{max}/10^{\frac{G}{20}}$. The threshold voltage $U_L$ is set to $U_H$–$U_F$, where $U_F$ represents the amplitude of the envelope detector's output. The thresholds $U_H$ and $U_L$ are tuned by two adjustable on-board resistors. In practice, considering that $A_{max}$ and $U_F$ both vary with the link distance, we measure these two values offline under different link distance settings and store a mapping table on each tag to facilitate the configuration of $U_H$ and $U_L$. To alleviate this manual configuration overhead, one could leverage an Automatic Gain Control (AGC) [42, 43] to adapt the power gain automatically. We leave it for future work.

### 4.2 LoRa Transmitter and Receiver

**LoRa transmitter**. We use two types of LoRa transmitters in the evaluation: *i*) a LoRa transmitter implemented on a software-defined radio platform USRP N210, and *ii*) a com-mercial off-the-shelf LoRa node equipped with a Semtech SX1276RF1JAS [7] chip. Both platforms use a single omni-directional antenna with 3 dBi gain. The transmission power is set to 20 dBm.

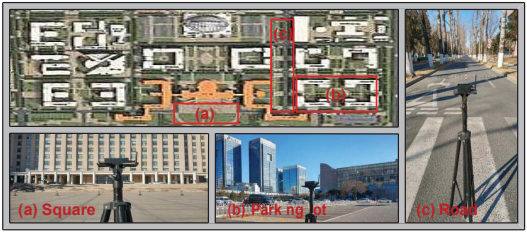

Figure 14: Outdoor experiment field.

**LoRa receiver**. The LoRa receiver is implemented on a software-defined radio platform USRP N210. We set the sampling rate to 10 MHz, thereby allowing the receiver to monitor six LoRa channels simultaneously.

### 4.3 ASIC Simulation

We simulate the Application Specific Integrated Circuit (ASIC) of Saiyan based on the TSMC 65-nm CMOS process. The active area of on-chip Integrated Circuits (IC) is 0.217 $mm^2$. The ASIC simulation shows that the power consumption of Saiyan is 93.2 $\mu W$. Specifically, the power consumption of LNA, oscillator, and digital circuit is 68.4 $\mu W$ and 22.8 $\mu W$, and 2 $\mu W$, respectively. Once Saiyan demodulated the feedback signals, the MCU starts preparing data for packet re-transmissions, which consumes extremely low power (*i.e.*, the power consumption of the ultra-low power Apollo2 [13] in Saiyan is merely 19.6 $\mu W$).

### 4.4 MAC-layer for Multi-tag Coexistence

We briefly discuss MAC-layer in this section. The downlink packets can be divided into three groups: unicast packet, multicast packet, and broadcast packet. In unicast, all backscatter tags within the radio range will receive and demodulate this unicast packet from the access point. However, only the targeted tag will response (*e.g.*, re-transmit the lost packet). Hence, no collision occurs. However, in multicast and broadcast, collision happens as long as more than one backscatter tag replies at the same time. For instance, the access point sends a downlink packet (*e.g.*, turn off the humidity sensor), while multiple tags acknowledge the reception of this downlink packet simultaneously. In this case, the access point can leverage slotted ALOHA [22] protocol to coordinate tags and minimize collisions. We take Figure 15 as an example to illustrate the MAC-layer operation. Suppose three tags are sending an acknowledgement to the access point to confirm the reception of a downlink packet. Each tag will randomly select a time slot and store it in its local counter. Upon the detection of a carrier signal from the access point, each tag decreases the slot number by one and transmits as soon as the slot number goes zero. The randomness in slot selection minimizes the interference among tags.

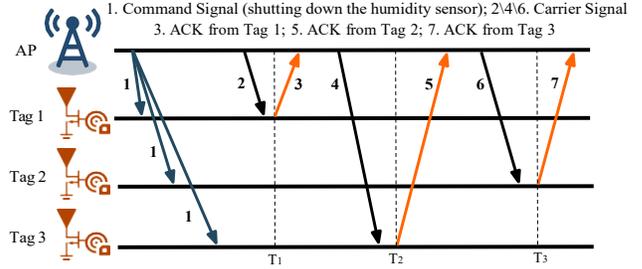

Figure 15: The illustration of MAC-layer operations in Saiyan. Each tag randomly selects a slot to transmit. The access point (AP) signals the beginning of each slot with a carrier signal.

## 5 Evaluation

In this section, we present the evaluation results of field studies (§5.1) and micro-benchmarks (§5.2). Two case studies follow (§5.3). Unless otherwise posted, the transmitter and the receiver are collocated throughout the experiment.

**Setups**. The LoRa transmitter works on the 433.5 MHz frequency band. The spreading factor and the bandwidth are set to 7 and 500 KHz, respectively. The payload of each LoRa packet contains 32 chirp symbols. In each experiment, we let the transmitter transmit 1,000 LoRa packets and then repeat the experiment for 100 times to ensure the statistical validity. We adopt *BER*, *throughput*, and *demodulation range* as the key metrics to assess Saiyan's performance.

- **BER** refers to the ratio of error bits to the total number of bits received by Saiyan.

- **Throughput** measures the amount of received data correctly decoded by Saiyan within one second.

- **Demodulation range** refers to the maximum distance between the tag and the LoRa transmitter when the BER is maintained below 1‰.

### 5.1 Field Studies

We conduct field studies both indoors and outdoors to assess the impact of coding rate (CR), spreading factor (SF), and bandwidth (BW) on BER, demodulation range, and throughput, which are three key evaluation metrics.

#### 5.1.1 Outdoor experiments

**Impact of coding rate**. We place a Saiyan tag 10 m, 20 m, 50 m, 100 m, and 150 m away from a LoRa transmitter. Under each distance setting, we vary the coding rate of LoRa signals and measure BER and throughput. We have three observations based on the results shown in Figure 16.

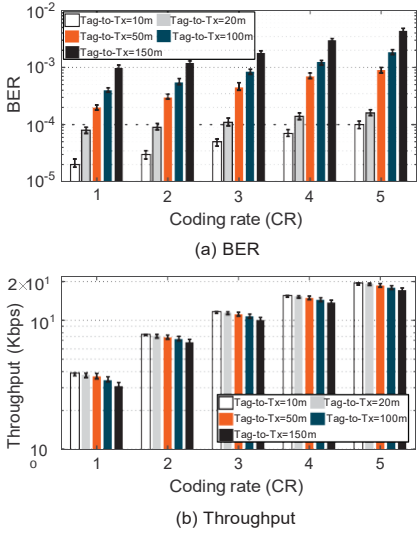
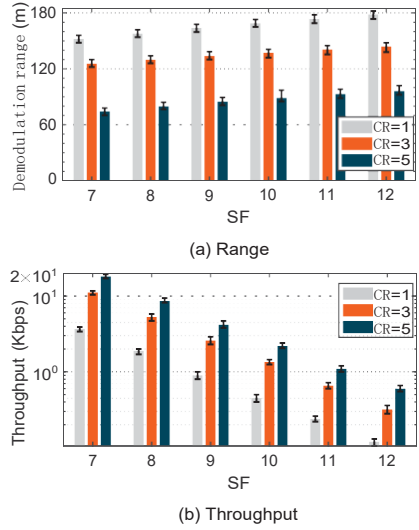
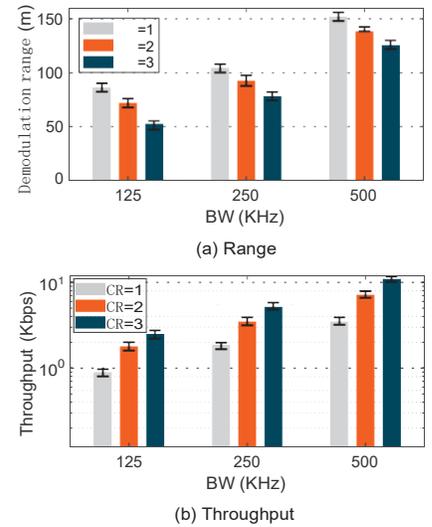

Figure 16: BER and throughput in different coding rate settings.

Figure 17: Demodulation range and throughput in different SF settings.

Figure 18: Demodulation range and throughput in different BW settings.

First, the BER grows with the coding rate. As shown in Figure 16(a), the BER under the highest coding rate setting (*i.e.*, 5) is 2.4–5.2× higher than the BER under the lowest coding rate setting (*i.e.*, 1) across all different Tx-to-tag distances. For instance, when the Tx-to-tag distance is 100 m, Saiyan achieves a BER of 1.85‰ under the highest coding rate setting. The BER then drops to 0.4‰ under the same Tx-to-tag distance setting when we change the coding rate to 1. This is expected since the Saiyan tag has to differentiate more types of LoRa chirps under the high coding rate setting.

Second, the throughput grows linearly with the coding rate (Figure 16(b)). For example, when the Tx-to-tag distance is 100 m, the achievable throughput at CR=5 (18.12 Kbps) is around 5.1× higher than the throughput at a coding rate of 1 (3.57 Kbps).

Third, both the BER and the throughput get exacerbated with the growing Tx-to-tag distance. For instance, when CR=5, the BER grows dramatically from 0.1‰ to 4.4‰ as the Tx-to-tag distance grows from 10 m to 150 m. The throughput, on the other hand, declines from 19.6 Kbps to 17.2 Kbps. This is expected since Saiyan relies on the signal power to demodulate the incident LoRa signal.

**Impact of spreading factor**. Next, we vary the spreading factor from 7 to 12 and assess Saiyan's demodulation range and throughput under each setting. The results are shown in Figure 17. We observe that the demodulation range grows with the increasing spreading factor. The throughput, on the contrary, declines with the increasing spreading factor. For instance, the demodulation range under the highest spreading factor setting (*i.e.*,SF=12) is 1.1–1.3× longer than the demodulation range under the lowest spreading factor setting (*i.e.*, SF=7) across three different coding rate settings. The throughput drops by 30.3–35.1× as we decrease the SF from 12 to 7. This is expected since a higher spreading factor enhances the anti-noise capability of LoRa signals; thus the demodulation range grows. On the other hand, the symbol time grows with the increasing spreading factor, resulting in a lower throughput.

**Impact of bandwidth**. We set the spreading factor to 7 and assess the impact of LoRa bandwidth on the demodulation range and throughput. The results are shown in Figure 18. We observe that the demodulation range and the throughput both grow with the LoRa bandwidth. Specifically, given the coding rate of 2, the demodulation range grows from 72.2 m to 138.6 m as we increase the bandwidth from 125 KHz to 500 KHz. On the other hand, since the LoRa symbol time is inversely proportional to the bandwidth, we observe the throughput drops around 4× from 7.2 Kbps to 1.8 Kbps as we decrease the bandwidth from 500 KHz to 125 KHz.

#### 5.1.2 Indoor experiments

We repeat the above experiments in an indoor environment where the LoRa signals have to penetrate one or multiple concrete walls to arrive at the backscatter tag.

**Penetrating one concrete wall**. Similar to the trend shown in the outdoor scenario, the throughput measured in the indoor scenario also grows with the increase of the coding rate (Figure 19). For example, the throughput grows from 3.7 Kbps to 18.7 Kbps when the coding rate varies from 1 to 5. The demodulation range, on the other hand, declines from 48.8 m to 26.2 m as we increase the coding rate from 1 to 5.

**Penetrating two concrete walls**. The LoRa signal experi-ences stronger attenuation when penetrating two concrete walls. Accordingly, we observe the demodulation range and the throughput decline by 2.21-2.09× and 1.01-1.05× compared to those under the single concrete wall settings (Figure 20).

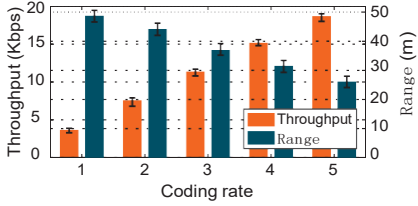
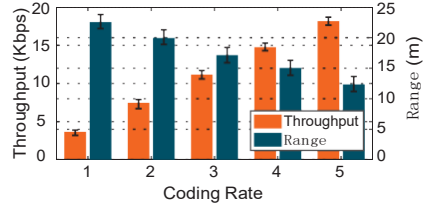
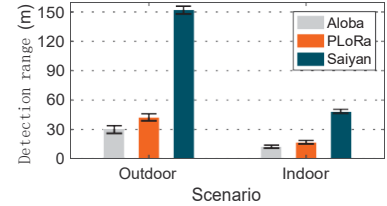

Figure 19: Throughput and downlink range in the presence of one concrete wall.

Figure 20: Throughput and downlink range in the presence of two concrete walls.

Figure 21: Comparison of Saiyan, Aloba, and PLoRa on the detection range.

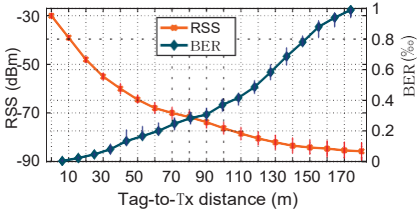
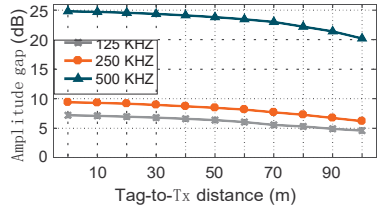
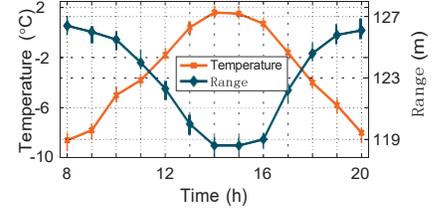

Figure 22: RSS and BER over distance.

Figure 23: The amplitude gap of the output signal after SAW filter

Figure 24: Demodulation range under different temperatures

### 5.1.3 Comparison with state-of-the-art systems

We further compare Saiyan with two state-of-the-art systems, namely, Aloba [23] and PLoRa [40] in both outdoor and indoor environments. PLoRa operates cross-correlation to detect a LoRa packet. Aloba feeds the incident signal into a moving average filter and then leverages the unique RSSI pattern of the LoRa preamble to detect a LoRa packet. They both cannot demodulate the payload. Therefore, we compare them with Saiyan in terms of the packet detection range.

Figure 21 shows the experiment result. In the outdoor line-of-sight settings, Saiyan achieves a packet detection range of 148.6 m, outperforming ALoBa (30.6m) and PLoRa (42.4m) by 4.52× and 3.26×, respectively. In an indoor none-line-of-sight environment, although the packet detection range of Saiyan declines to 44.2 m, it still outperforms Aloba (12.4 m) and PLoRa (16.8 m) by 3.56× and 2.63×, respectively.

## 5.2 Micro-benchmarks

To better understand the performance of each design component in Saiyan, we run micro-benchmarks to assess the receiver sensitivity, the SAW filter, as well as the power consumption and the system cost.

### 5.2.1 Receiver sensitivity

We define the receiver sensitivity as the minimum Received Signal Strength (RSS) of an incident signal that can be detected by Saiyan. To assess the receiver sensitivity, we measure the BER and the Received Signal Strength (RSS) under different Tx-to-tag distance settings. As expected, the BER grows gradually with the increase of the Tx-to-tag distance, as shown in Figure 22. Nevertheless, Saiyan can still detect the incident signal when the tag is 180 m away from the transmitter. As we increase the tag-to-Tx further, the signal strength is too weak to be detected by Saiyan. The above experiment demonstrates an -85.8 dBm receiver sensitivity, outperforming the conventional envelope detector by 30 dBm [27].

### 5.2.2 Performance of the SAW filter

**Frequency-amplitude response**. Saiyan relies on the frequency-amplitude response of the SAW filter to demodulate LoRa signals. A sharp frequency-amplitude response (*e.g.*, a small frequency variation leads to a large amplitude gap) is desirable as it allows the Saiyan tag to detect the minute frequency variation on the incident signal.

We feed LoRa signals with different bandwidth into the SAW filter and measure the amplitude variation of the output signal. The results are shown in Figure 23. As expected, the amplitude variation of the output signal (*a.k.a.*, amplitude gap) tends to be less significant with the decreasing chirp bandwidth. For instance, when the Tx-to-tag distance is 10 m, the amplitude gap drops from 24.7 dBm to 9.3 dBm, and further to 7.1 dBm as we decrease the chirp bandwidth from 500 KHz to 250 KHz, and further to 125 KHz, respectively. A similar trend shows up as we increase the Tx-to-tag distance.

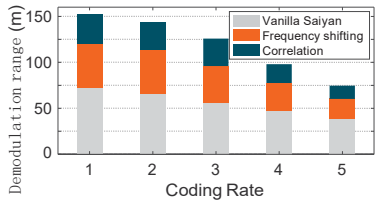

Figure 25: Ablation study of Saiyan.

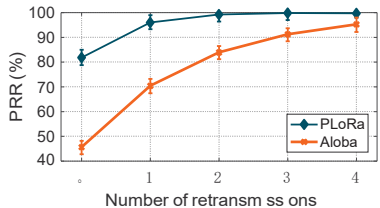

Figure 26: PRR in different settings.

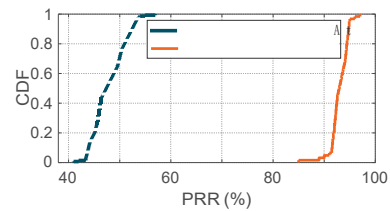

Figure 27: PRR lifts with Saiyan.

Table 2: Energy consumption (under 1% duty cycling) and cost of each component in Saiyan tag.

| Component | SAW Filte | LN | OSC Clock | Envelope Detector | Comparator | MCU | Total |
|---|---|---|---|---|---|---|---|
| Energy ($\mu$W) | 0 | 248.5 | 86.8 | 0 | 14.45 | 19.6 | 369.4 |
| Cost ($) | 3.87 | 4.15 | 1.25 | 1.20 | 1.26 | 15.43 | 27.2 |

For instance, when the signal bandwidth is 500 KHz, the amplitude gap of the output signal drops from 24.7 dBm to 20.2 dBm as the Tx-to-tag distance increases from 10 m to 100 m.

**The impact of temperature**. The frequency selectivity of the SAW filter is affected by the ambient temperature [36]. We thus run an experiment to assess the impact of temperature on the demodulation range. The experiment is conducted outdoors on a sunny day from 8 a.m. to 8 p.m.. Figure 24 shows the result. We observe that the demodulation range in general is insensitive to the temperature. For instance, when the temperature rises from the lowest -8.6 °C at 8 a.m. to the highest 1.6 °C at 2 p.m., the demodulation range merely drops from 126.4 m to 118.6 m.

### 5.2.3 Ablation study

We conduct an ablation study to assess the effectiveness of each design component of Saiyan. In this experiment, we set the spreading factor and the bandwidth to 7 and 500 KHz respectively and measure the maximum demodulation range under different coding rate settings. The results are shown in Figure 25. We find that the vanilla Saiyan achieves a relatively short demodulation range (38.4 m—72.6 m) across five different coding rate settings. The demodulation range then grows by 1.56×–1.73× with the help of the cyclic frequency shifting module. The cross-correlation further improves the demodulation range by 1.94×–2.25×.

### 5.2.4 Power consumption & system cost

Table 2 summarizes the power consumption (under 1% duty cycling as in LoRa [22]) and cost of each component in Saiyan. Among these hardware components, the most power-hungry parts are LNA and oscillator (OSC) clock, which account for 67.3% and 23.5% of the total power consumption, respectively. As we demonstrate in §4.3, the power consumption can be effectively reduced by 74.8% when implementing Saiyan on ASIC. The hardware cost of Saiyan, on the other hand, is around 27.2 USD, which can be also reduced sharply after ASIC fabrication.

## 5.3 Case Studies

Next, we run two real-world case studies to showcase packet re-transmission (§5.3.1) and frequency hopping (§5.3.2).

### 5.3.1 Packet re-transmission through the ACK mechanism

**Setups**. We integrate Saiyan into PLoRa and Aloba tags, which allows the tags to demodulate the feedback signals from the receiver and make an immediate packet re-transmission if needed. The link distance is set to 100 m.

**Results**. As shown in Figure 26, PLoRa and Aloba achieve 81.8% and 45.6% packet reception ratio (PRR) without packet re-transmission. The PRR of Aloba grows drastically from 45.6% to 70.1% when the Aloba tag is allowed to re-transmit the lost packet only once. The PRR then grows to 83.3$ and further to 95.5% when the Aloba tag re-transmits the lost packet twice and three times, respectively. The PRR of PLoRa shows the similar trend. These results demonstrate that Saiyan effectively improves the packet reception ratio for long-range LoRa backscatter systems.

### 5.3.2 Interference avoidance through channel hopping

As an ultra-low-power tag working on the ISM band, both PLoRa and Aloba are likely to bear strong in-band interference from other legacy RF devices working on the same band. We show that with Saiyan, these backscatter tags can demodulate the feedback signals from the receiver and switch to other channels to avoid interference.

**Setups**. We use PLoRa to demonstrate the feasibility of channel hopping. The PLoRa tag communicates with the receiver at the 434 MHz frequency band. It switches to the 434.5 MHz frequency band upon detecting the feedback signal from the receiver. We put a software-defined radio three meters away from the receiver to jam the channel at the 433 MHz frequency band.

**Results**. Figure 27 shows the CDF of PRR before and after the channel hopping. We can see the PRR is very low when the USRP jams the channel (dotted line). As the receiver initiates a channel hopping command to the backscatter tag, we witness a significant lift on the PRR. In particular, the median PRR grows from 47% to 92% once PLoRa switches to another channel. This result clearly demonstrates that Saiyan can support better channel utilization through remote control.

## 6 Related Work

We review research topics relevant to Saiyan in this section.

**RFID system.** A passive RFID tag modulates sinusoidal tone from an RFID reader to transmit data [52, 58]. It can also demodulate amplitude-modulated (AM) signals from a nearby RFID reader [16, 26, 51, 53]. Specifically, the RFID tag downconverts the incident signal to the baseband and accumulates the signal power through an integrator circuit. Subsequently, it compares the accumulated power to a threshold to demodulate incident signals. Saiyan differs from passive RFID tags in two aspects. First, Saiyan demodulates frequency-modulated signal as opposed to amplitude-modulated signal. Second, Saiyan is designed for long-range backscatter systems whereas the passive RFID tag functions within only a few meters.

**Ambient backscatter systems.** Ambient backscatter systems empower backscatter tags to take the ambient wireless traffic as the carrier signals [14, 15, 18, 20, 23, 29–33, 35, 37, 39, 40, 47–50, 55–57, 60]. For example, WiFi backscatter [33] reuses WiFi signals as the carrier, thereby allowing for the backsactter tag to communicate with a commercial WiFi receiver. Interscatter [29] enables backscatter tags to modulate Bluetooth signals into WiFi signals. LoRa backscatter [47] allows backscatter tags to communicate over long distances by taking advantage of the noise resilience of LoRa symbols. These pioneer works have remarkably improved the throughput and the communication range of backscatter systems. Some recent works [37, 44, 55, 56, 59, 60] support a few types of downlink functionalities such as carrier sensing [37, 44, 55, 56, 59, 60] and packet detection [23, 40] at the packet level. For example, WiFi backscatter [33], Passive-WiFi [34], Interscatter [29], LoRa backscatter [47], and Netscatter [24] use the presence and absence of carrier packets to convey downlink data. However, they cannot demodulate downlink packets at the symbol level, particularly under long-range settings. Saiyan can serve as an important building block to the existing long-range backscatter systems, where the on-demand retransmission is needed due to the drastic packet loss.

**Low-power demodulator.** With the growth of low-power IoT market, the research community has shifted the focus to the design and implementation of low-power RF receivers, *e.g.*, by replacing the active components with their passive counterparts, or by offloading the power-intensive functions to external devices. Ensworth et al. [19] proposed a 2.4 GHz low-power BLE receiver that offloads the RF local oscillator to an external device. Carlos et al. [41] proposed a low-power 802.15.4 receiver that could demodulate phase-modulated ZigBee signals at orders of magnitude lower power consumption compared with the standard 802.15.4 receiver. However, the working range of this low-power receiver is limited to tens of centimeters, which sets a strong barrier towards the practical deployment. Turbo charging [39] designs a multi-antenna cancellation circuit to facilitate the signal demodulation on backscatter tags. Similarly, full-duplex backscatter [38] enables a backscatter tag to demodulate the instantaneous feedback signal from another backscatter tag. Saiyan differs from these systems in two aspects. First, Saiyan is designed for demodulating frequency-modulated signals as opposed to phase or amplitude modulated signals. Second, Saiyan can support up to 180 m demodulation range, whereas all the aforementioned systems function within only tens of centimeters.

**SAW filter.** The SAW filter has been widely adopted by wireless communication systems such as telecommunications [25], radar [54], and aerospace communications [45], *etc.* These systems leverage the low-distortion and minimal passband variation of the SAW filter to filter out noise and interference signals. Furthermore, medical devices transform a SAW filter into a sensor for in-situ detection (*e.g.*, detecting chemical gas concentration) [21, 28]. Different from all the above applications, Saiyan exploits the sharp frequency response of the SAW filter to demodulate frequency-modulated signal.

## 7 Conclusion

We have presented the design, implementation, and evaluation of Saiyan, the first-of-its-kind low-power demodulator for LoRa backscatter systems. Saiyan allows LoRa backscatter tags to demodulate the command or feedback signals from a remote access point that is hundreds of meters away. With such capability, the backscatter tag can realize a plethora of networking functionalities, such as packet re-transmission, channel hopping, and rate adaptation. Field study shows that Saiyan outperforms state-of-the-art systems by 3.5–5 in terms of demodulation range. The ASIC simulation shows that the power consumption of Saiyan is around 93.2 $\mu W$.

## Acknowledgment

We thank our shepherd Fadel Adib and the anonymous reviewers for their insightful comments. We are also very grateful to Dr. Lu Li from University of Electronic Science and Technology of China for his constructive feedback. This work is supported in part by National Key R&D Program of China No. 2017YFB1003000, National Science Fund of China under grant No. 61772306, and the R&D Project of Key Core Technology and Generic Technology in Shanxi Province (2020XXX007).

## A Appendix

In this section, we prove the infeasibility of RLC resonant circuit to realize LoRa frequency-amplitude transformation.

### A.1 The Infeasibility of RLC Resonant Circuit

The center frequency $\omega_0$, the passband $\Delta\omega$, and the quality factor $Q$ of a resonant circuit satisfy that:

$$Q = \frac{\omega_0}{\Delta\omega} \quad (6)$$

A higher $Q$ value leads to a narrower passband width. Taking a step further, the quality factor $Q$ is determined by the resistance $R$, inductance $L$, and capacitance $C$ of this circuit following the equation:

$$Q = \sqrt{L}/(R \cdot \sqrt{C}) \quad (7)$$

Given a constant center frequency of $\omega_0 = 1/(2\pi\sqrt{LC})$, we can deduce the capacitance $C$ satisfy that:

$$C = \frac{1}{Q\omega_0 R} \quad \frac{\Delta\omega}{\omega_0^2 R} \quad (8)$$

Generally, the equivalent $R$ of RF circuit is 50 Ω. Taking LoRa signals working on 433 MHz frequency band (with 500 KHz bandwidth) as an example, this requires $C$ to be as low as $5.2 \times 10^{-14} pF$.